\documentclass[aps,superscriptaddress,prb,twocolumn]{revtex4-1}
\usepackage{graphicx,bm,color,amssymb,amsmath,cancel}

\usepackage{dcolumn}
\usepackage{subfigure}
\usepackage{siunitx}
\usepackage{multirow}
\usepackage[english]{babel}
\usepackage[colorinlistoftodos]{todonotes}
\usetikzlibrary{fit,matrix}
\usepackage{bbold}
\usepackage[english]{babel}
\usepackage{verbatim}
\usepackage{appendix}
\usepackage[normalem]{ulem}

\newcommand{\zqi}[1]{\textcolor{black}{#1}}
\newcommand{\ket}[1]{|{#1}\rangle}
\newcommand{\bra}[1]{\langle{#1}|}

\def\simlt{\mathrel{\lower .3ex \rlap{$\sim$}\raise .5ex \hbox{$<$}}}

\DeclareMathOperator{\dif}{d\!}

\begin{document}
	
	
	\title{Effects of charge noise on a pulse-gated singlet-triplet $S-T_-$ qubit }
	
	\author{Zhenyi Qi}
	\author{X. Wu}
	\affiliation{University of Wisconsin-Madison, Madison, WI 53706}
	\author{D. R. Ward}
	\affiliation{University of Wisconsin-Madison, Madison, WI 53706}
	\affiliation{Center for Computing Research, Sandia National Laboratories, Albuquerque, NM 87185, USA }
	\author{J. R. Prance}
	\author{Dohun Kim}
	\affiliation{University of Wisconsin-Madison, Madison, WI 53706}
	\author{John King Gamble}
	\affiliation{University of Wisconsin-Madison, Madison, WI 53706}
	\affiliation{Center for Computing Research, Sandia National Laboratories, Albuquerque, NM 87185, USA }
	\author{R. T. Mohr}
	\author{Zhan Shi}
	\author{D. E. Savage}
	\author{M. G. Lagally}
	\author{M. A. Eriksson}
	\author{Mark Friesen}
	\author{S. N. Coppersmith}
	\author{M. G. Vavilov}
	\affiliation{University of Wisconsin-Madison, Madison, WI 53706}
	
	\date{\today}
	
	\begin{abstract}
		We study the dynamics of a pulse-gated semiconductor double quantum dot qubit.  In our experiments, the qubit coherence times are relatively long, but the visibility of the quantum oscillations is low. We show
		that these observations are consistent with a theory that incorporates decoherence arising from charge noise that gives rise to detuning fluctuations of the double dot. Because effects from charge noise are largest near the singlet-triplet avoided level crossing, the visibility of the oscillations are low when the singlet-triplet avoided level crossing occurs in the vicinity of the charge degeneracy point crossed during the manipulation, but there is only modest dephasing at the large detuning value at which the quantum phase accumulates.
		This theory agrees well with experimental data and predicts that the visibility can be increased greatly by appropriate tuning of the interdot tunneling rate.
	\end{abstract}
	
	\maketitle
	
	
	\section{Introduction}
	Electrically-gated solid-state qubits fabricated using quantum dots in semiconductors are attractive because of the similarity of the technology to that used in current classical electronic devices, with the great potential advantages of scalability and relative ease of qubit manipulation~\cite{Loss1998,Levy2002,Vrijen2000,Awschalom1174,Shulman2014}. 
	Quantum dot qubits in gallium arsenide (GaAs) heterostructures~\cite{Petta2005, Koppens2006,Reilly817, Foletti2009, Barthel2009, Barthel2010,Bluhm2011, Nowack2011, Gaudreau2012, Shulman2012, Nichol1608}
	in the absence of dynamic nuclear polarization display fast dephasing (on nanosecond time scales) due to the strong hyperfine interaction between electron and nuclear 
	spins~\cite{Johnson2005,Reilly2008, Barthel2009, Barthel2010}.
	Electrons in silicon quantum dots have weaker coupling to nuclear spins~\cite{Dyakonov1992, Assali2011}, and measured
	qubit coherence times are indeed longer,
	on the order of several hundred nanoseconds\ \cite{Maune2012, Wu2014, Tyryshkin2005, Tyryshkin2011, Takeda2016} for natural silicon and even longer for isotopically enriched silicon\cite{Eng1500214, Veldhorst2014, Muhonen2014}.
	Integrating a micromagnet into a double quantum dot device enables the establishment of a large magnetic field difference between the dots that does not depend on the presence of nuclear spins~\cite{Pioro2007, Pioro-Ladriere2008, Obata2010,Wu2014, Kawakami2014,Takeda2016, Kawakami2016, Scarlino2015}, 
	enabling fast spin manipulations without introducing a magnetic source of decoherence.
	
	In this paper we study Landau-Zener-St\"{u}ckelberg (LZS) oscillations that are
	performed by pulsing through an $S-T_\pm$ anticrossing in a double quantum dot fabricated
	in a silicon/silicon-germanium (Si/SiGe) heterostructure with an integrated micromagnet.
	LZS oscillations were demonstrated first in GaAs devices \cite{Barthel2009, Petta2005, Petta2010, Ribeiro2010, Stehlik2012, Ribeiro2013a, Ribeiro2013, Cao2013}.
	In the GaAs experiments, the coherence time of the LZS oscillations is short, $\sim$10 ns,
	with an oscillation visibility of $\sim$30\%~\cite{ Petta2005, Barthel2009, Ribeiro2013}.
	We report LZS experiments performed in a Si/SiGe heterostructure for a variety of ramp rates
	and find that the decoherence times are indeed much longer,  $\sim 1.7{~\rm \mu s}$, but that the visibility of the qubit
	oscillations is only $\simlt$30\%.
	We then demonstrate that these observations can be understood as a consequence
	of the presence of charge noise.
	Dephasing from charge noise has been argued previously to be important
	for LZS experiments~\cite{Ribeiro2013a, Dial2013, Taylor2007, Nichol2015}, 
	and numerical simulations have yielded strong evidence that charge noise effects are substantial~\cite{Nichol2015, SINichol2015}.
	Here we argue that because the energy splitting at the relevant anticrossing is much smaller than
	the temperature, excitations across the energy gap play a critical role.
	The effects of charge noise are substantial only near the charge transition and are much smaller at large
	detunings where the spin rotations are performed,
	so the measured spin coherence times can be long even though the visibility is low.
	Our theoretical treatment yields analytic insight into the processes limiting the oscillation visibility.
	We show that the visibility can be increased substantially by changing the dot parameters, specifically,
	by increasing the interdot tunnel coupling.
	
	\section{Landau-Zener-St\"{u}ckelberg Interferometric Measurements}
\begin{figure}
		\begin{minipage}{0.2\textwidth}
			\subfigure{%
				\includegraphics[scale = 0.3]{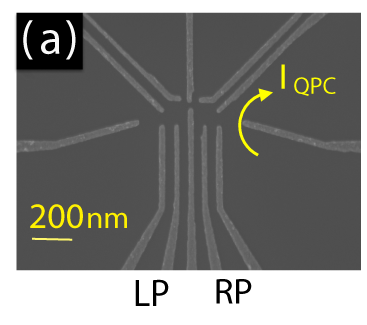}
				\label{fig:deviceDiagram}}
		\end{minipage}~
		\begin{minipage}{0.22\textwidth}
			\subfigure{%
				\includegraphics[scale = 0.3]{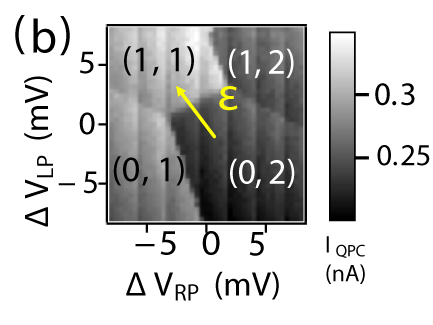}
				\label{fig:stabilityDiagram}}
		\end{minipage}\hfill
		\caption{(Color online)
			(a) Micrograph of a double-dot device in a Si/SiGe heterostructure that is lithographically identical to the one used in the experiment~\cite{Wu2014}. 
			(b) Stability diagram showing the electron occupations in the dots, obtained by measuring the current through the quantum point contact, $I_{\mathrm{QPC}}$, at different voltages $\Delta V_{\mathrm{LP/RP}}$ applied to the left/right gates labeled in (a).
			The numbers in parentheses are the electron occupations of the two dots.
		}
	\end{figure}

	A micrograph of a Si-based double quantum dot that is identical to the device in the experiment is shown in Fig.~\ref{fig:deviceDiagram}. By measuring the current through the quantum point contact (QPC), indicated by the yellow arrow, the charge occupation of each dot can be determined, as shown in the charge stability diagram in Fig.~\ref{fig:stabilityDiagram}. The number of electrons on each dot is shown on the diagram. 
	
\begin{figure}[h]
		\begin{minipage}{0.48\textwidth}
			\subfigure{%
			\includegraphics[scale = 0.3]{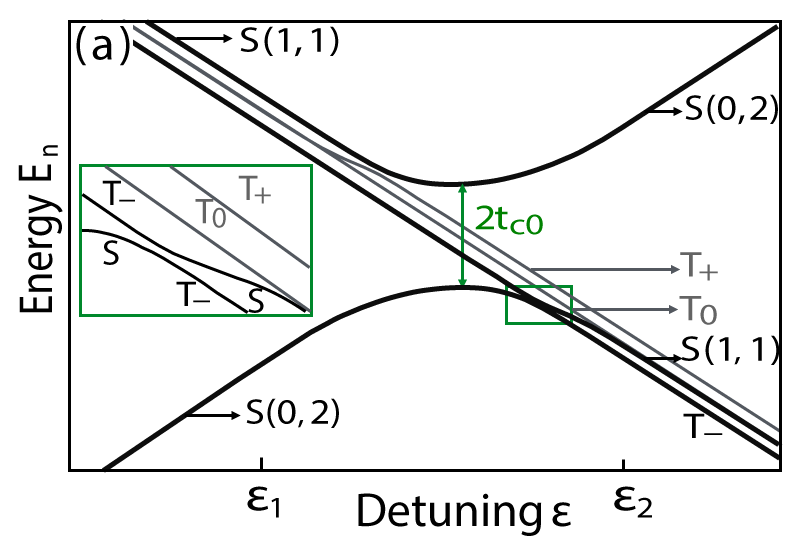}
			\label{fig:energyDiagram_sub}}
		\end{minipage}
		\begin{minipage}{0.36\textheight}
			\subfigure{%
			\includegraphics[scale = 0.3]{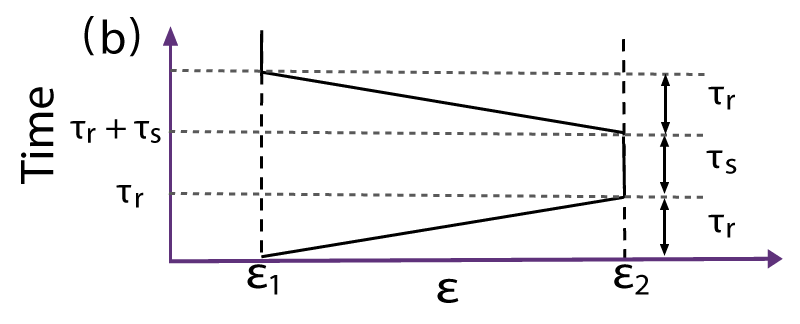}
			\label{fig:pulse_shape}}
		\end{minipage}
		\caption{(a) Schematic energy diagram of the full five-level system. A small transverse magnetic field gradient causes an anticrossing to occur between the singlet and triplet states $S(1,1)$ and $T_-$. The left inset is an expanded view of the region in the small green box in the main figure. (b) a schematic of the pulse applied to the detuning $\varepsilon$ as a function of time $t$. The system is ramped from a negative detuning $\varepsilon_1$ to a large positive detuning $\varepsilon_2$ over a ramp time $\tau_r$, held at $\varepsilon_2$ for a manipulation time $\tau_s$, and then ramped back to $\varepsilon_1$ over the time $\tau_r$. The pulse sequence passes through the $S-T_-$ anticrossing twice, giving rise to Landau-Zener-St\"{u}ckelberg oscillations. }
		\label{fig:energyDiagram}
	\end{figure}

	The device is {fabricated} with a micromagnet that induces a magnetic field difference between the dots, $\delta B$, and also a uniform magnetic field that, combined with an external magnetic field plus the magnetic fields from nuclear spins, gives rise to a Zeeman splitting between the triplet states. The transverse component of $\delta B$ induces an anticrossing between the singlet state $\ket{S}$ and spin-polarized triplet $\ket{T_-}$. 
	Fig.~\ref{fig:energyDiagram_sub} shows the schematic energy diagram of the double quantum dot along the detuning direction, indicated by the yellow arrow in Fig.~\ref{fig:stabilityDiagram}. The left inset is a blowup of the region near the $S-T_{-}$ anticrossing.
	
	The pulse sequence used in the experiment is shown in Fig.~\ref{fig:pulse_shape}.  The detuning is ramped from a negative value through the $S-T_-$ anticrossing to a large positive value, where it is held for a manipulation time $\tau_s$, and then it is ramped back to the initial value, where it is held long enough for the spin state to be measured and reset. When the ramp rate is appropriate, the first ramp leads to occupation of both states with a relative phase that accumulates at large detuning during the manipulation time, and ramping back to (2,0) gives rise to Landau-Zener-St\"{u}ckelberg (LZS) oscillations. The probability of being in the singlet state at the end of the sequence oscillates as a function of $\tau_s$, as shown in the inset of Fig.~\ref{fig:visiRate} by a red solid line. These data were taken with a ramp time of $\tau_r \simeq {45}{~\rm ns}$, which corresponds to a ramp rate of $\sim$4.4$~\mu$eV/ns. 
	The coherence time extracted from the oscillations is quite long ($\sim1.7~{\rm \mu s}$), but the visibility, defined as the maximum amplitude of the oscillations, is only about $0.24$, much less than the value of $1$ expected for LZS oscillations in the absence of decoherence~\cite{landau1932theorie,stuckelberg1932theory,Shevchenko2010}. In the experiment, the ramp rate was varied, yielding a peak in the visibility, as indicated by the red circles in the main panel of Fig.~\ref{fig:visiRate}. Here, the data in the inset correspond to a point near the top of the peak.
	
	\section{Model}
	Since we are mainly interested in the singlet-triplet $S-T_-$ subspace, we first reduce the full five-level system to two levels, as described in Appendix. The resulting qubit Hamiltonian can be written as a $2\times 2$ matrix:
	
	\begin{equation}
	\hat{H}_{S-T_-}^{(1)} = \begin{pmatrix}
	-E_S &  \dfrac{h}{2} \\
	\dfrac{h}{2} & E_{T_-} 
	\end{pmatrix},
	\label{eq:H1st}
	\end{equation}
	where the singlet and triplet energies are given by $E_S = \sqrt{(\varepsilon/2)^2+t_c^2}$ and $E_{T_-} = -(\varepsilon/2 +g\mu_BB)$. As in Ref. [22], we assume that the tunnel coupling between the quantum dots, $t_c=t_{c0}\exp(-\varepsilon/\varepsilon_0)$, varies exponentially with the detuning $\varepsilon$, and $\varepsilon_0 \simeq 125 ~{\rm \mu eV}$. Here, $g$ is the gyromagnetic ratio, $mu_B$ is the Bohr magneton, $B$ is the average total magnetic field on the two dots, and the detuning is defined such that $\varepsilon=0$ at the charge degeneracy point. In Eq.~\eqref{eq:H1st}, the transverse magnetic field gradient causes hybridization of the $S$ and $T_-$ states through the parameter $h=\sqrt{2}h_x\cos(\theta/2)$, where $h_x=g \mu_B \delta B_x$ and $\theta=\arccos(\varepsilon/2E_S)$; see Appendix for details. 
	
	We now apply a unitary transformation  
	\begin{equation}
		\hat{U} =  \exp( i \hat{\sigma}_y \phi/ 2),\quad
		\phi=\arccos\left(-\frac{E_{T_-}+E_S}{\Delta}\right).
		\label{eq:U2}
	\end{equation} 
	with $\Delta=\sqrt{(-E_S-E_{T_-})^2+h^2}$, to diagonalize the instantaneous Hamiltonian, Eq.~\eqref{eq:H1st}, and obtain
	
	\begin{equation}
	\begin{aligned}
		\hat H_{S-T_-}^{(2)} 
		& = \hat U \hat H_{S-T_-}^{(1)}  \hat{U}^\dagger - i\hat U  
		\dot{\hat{U}}^\dagger\\
		& = \frac{E_{T_-}-E_S}{2}\hat 1_{2\times 2}+\frac{\Delta}{2}\hat{\sigma}_z-\frac{\dot \phi}{2}\hat{\sigma}_y.
		\label{eqn:H2st}
	\end{aligned}
	\end{equation}

	Here, $\hat{1}_{2\times 2}$ is the two-dimensional identity matrix, and $\hat{\sigma}_{y,z}$ are Pauli matrices. Note that $E_{T_-}$ and $E_S$ are functions of time during the ramping pulse, resulting in the time dependence of $\phi$.
	There are two major noise sources in the double quantum dots: nuclear magnetic field fluctuations~\cite{Petta2010, Ribeiro2010, Ribeiro2013a, Ribeiro2013, Wong2015, Castelano2016} and charge noise~\cite{burkard1999coupled, Nichol2015, Wang2013, thorgrimsson2016}. Here, 
	We disregard the nuclear magnetic field fluctuations, based on arguments given below. Charge noise is included by incorporating fluctuations in the detuning $\varepsilon$~\cite{Dial2013}, so that $\varepsilon\to \varepsilon +\delta \varepsilon$ is a sum of a controlled gate detuning $\varepsilon$ and a fluctuating component $\delta \varepsilon$.
	In the original five-dimensional basis, the noise takes the diagonal form $\hat{V} ^{(0)} = (\delta\varepsilon/2) {\rm diag}(1, -1, -1, -1, -1)$. After applying the transformation $\hat{W}$ given by Eq.~\eqref{eq:U1}, in the qubit subspace, the noise contribution to the Hamiltonian takes the form $\hat{V} ^{(1)} = -(\delta\varepsilon/2){\rm diag}(\cos\theta, 1)$. We then apply transformation $\hat{U}$, Eq.~\eqref{eq:U2}, obtaining
	\begin{equation}
		\hat V^{(2)}=\frac{\delta \varepsilon}{2}
		(\sin\phi\hat{\sigma}_x+\cos\phi \hat \sigma_z )\sin^2\frac{\theta}{2},
		\label{eqn:V2}
	\end{equation}
	omitting a term proportional to the identity. The $\hat{\sigma}_x$ term in Eq.~\eqref{eqn:V2}, which causes transitions between the ground and the first excited states, affects the evolution of the density matrix during the ramp across the magnetic anticrossing. The $\hat{\sigma}_z$ term in Eq.~\eqref{eqn:V2} is diagonal but vanishes at the magnetic anticrossing when $\phi=\pi/2$; This term gives rise to fluctuations of the phase difference between the ground and first excited states. Below, we discuss the effects of these two terms on the system dynamics, \zqi{showing that the former explains the low visibility, while the latter give rise to dephasing that is consistent with the experiment. In this way, we obtain a self-consistent description of the experimental data in Figs.~\ref{fig:visiRate} and \ref{fig:PsTau_exp} }.

	We now apply the Bloch-Redfield (BR) approximation~\cite{Bloch1957, Redfield1957, Xu2014} to describe the dynamics of the double quantum dot in the presence of detuning noise $\hat{V}^{(2)}$. Within this theory, the dynamics are described in terms of transition rates between energy eigenstates in the $S-T_-$ subspace. 
	The detuning fluctuations are characterized by the spectral function
	$S(\omega)=\int d\tau \langle \delta \varepsilon(t) \delta \varepsilon(t+\tau)\rangle e^{-i\omega \tau}$, where $\omega$ is the frequency and $\langle \dots \rangle$ denotes an average over noise realizations. Here, we assume that the noise spectrum for detuning fluctuations has spectral density 
	\begin{equation}
		S(\omega) = \alpha/\omega^{0.7}, 
	\end{equation}
	where $\alpha$ is a constant, as consistent with experiment measurements on recent experimental quantum dot qubits~\cite{Dial2013}. We then calculate transition rates between eigenstates induced by the noise using Fermi's golden rule, obtaining the master equations for the qubit density matrix $\rho$ describing the hybridized $S-T_-$ two-level system: 
	\begin{subequations}
		\begin{align}
		\dot{\rho}_{00} & = \frac{\dot{\phi}}{2}(\rho_{01}+\rho_{10}) -\Gamma \rho_{00} + \Gamma \rho_{11} 
		\\ \dot{\rho}_{01} & = -\frac{\dot{\phi}}{2}(\rho_{00}-\rho_{11}) - \frac{i}{\hbar}\rho_{01}(\Delta + Z\delta\varepsilon)-\Gamma \rho_{01} \label{subeq:off-diagonalrho}
		\\ \dot{\rho}_{11} & = -\frac{\dot{\phi}}{2}(\rho_{01}+\rho_{10})+\Gamma \rho_{00} - \Gamma \rho_{11}~,
		\end{align}
		\label{eqn:differentialequations}
	\end{subequations}
	and $\rho_{10}=\rho_{01}^*$, where $Z = \sin ^2(\theta/2) \cos(\phi)$, and $Z\delta\varepsilon$ is the noise contribution from the $\hat{\sigma}_z$ term in Eq.~\eqref{eqn:V2}. Here, $0$ and $1$ refer to the instantaneous eigenstates of Eq.~\eqref{eq:H1st}.
	
	The transition rate $\Gamma$ in Eq.~\eqref{eqn:differentialequations} characterizes the rate of excitation from the ground state to the excited state and the rate of relaxation of the excited state. To estimate $\Gamma$, \zqi{we assume the delta epsilon noise distribution to be classical because the transition rates between $S$ and $T_-$ are only non-negligible near the $S-T_-$ anticrossing, where the energy separation between the levels is $~h$, which is very small compared to the temperature. In the BR theory we therefore obtain the following form for the relaxation rate:}
	
	\begin{equation}
		\Gamma(\varepsilon) =  \dfrac{\pi}{2\hbar^2}\sin^2\phi\sin^4\left(\dfrac{\theta}{2}\right)S (\Delta(\varepsilon) /\hbar )~.\label{eqn:transRate}
	\end{equation}
	\zqi{Charge noise and other sources of relaxation give rise to $T_1 \sim 10~ {\rm \mu s} $, as measured in experiments.\cite{Wu2014}  Since $T_1$ is much larger than the ramp time ($10\sim 100 ~{\rm ns}$) in our experiment, we ignore it in our visibility calculations.} 

	\section{Results}
	\subsection{Oscillation Visibility}
	\begin{figure}
		\includegraphics[scale = 0.3]{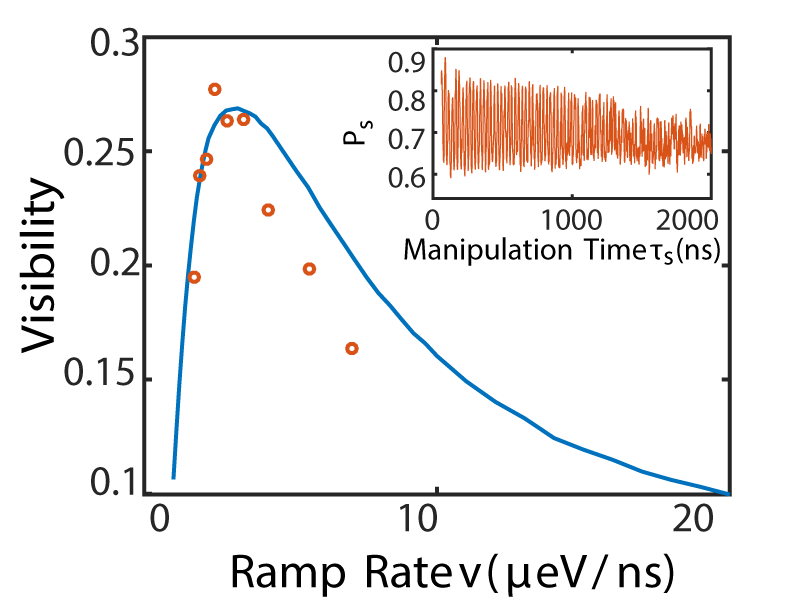}
		\caption{(Color online) Visibility of LZS oscillations as a function of ramp rate $v$, where $v$ is defined as the slope $\mathrm{d} \varepsilon/\mathrm{d} t$ of the initial ramp in the pulse sequence of Fig.~\ref{fig:pulse_shape}. The red dots are experimental data, and the blue line shows the results of theoretical simulations incorporating charge noise with a $\alpha/\omega^{0.7}$ spectrum~\cite{Dial2013}. No magnetic fluctuations are included in the model. The tunnel coupling \zqi{at zero detuning} used in the calculation is the same as measured in the experiment, $t_{c0} = {3.4}{~\rm \mu eV}$. \zqi{The adjustable parameters used to obtain the theoretical results are $\alpha$, which determines the noise amplitude and yields detuning fluctuations consistent with experimental estimates~\cite{Wu2014}, and $h$, which describes the transverse magnetic field gradient and determines the optimal ramp rate.} The inset shows the experimental return probability $P_s$ measured as a function of manipulation time $\tau_s$. }
		\label{fig:visiRate}
	\end{figure}

	We now compare the results of our numerical simulations of the differential Eq.~\eqref{eqn:differentialequations} to experimental measurements of Landau-Zener-St\"{u}ckelberg interferometry.
	The experiments presented here use the procedures and methods presented in Ref.~[\onlinecite{Wu2014}].
	The simulations use the measured values for the average magnetic field (obtained from the period
	of the LZS oscillations), $g\mu_B B = {0.17~}{\rm \mu eV}$, and the tunnel coupling \zqi{at zero detuning $t_{c0} \approx {3.4~}{\rm \mu eV}$~\cite{Wu2014} },
	which determines the value of the detuning at which the $S-T_-$ anticrossing occurs as a function of applied magnetic field.
	We ignore the term $Z\delta\varepsilon$ in Eq.~\eqref{subeq:off-diagonalrho} during the forward and backward ramps because it vanishes at $\phi = \pi/2$ and is only important for long times, as we explain below, in Sec.~\ref{subsec:dephasing}.
	The parameters $h$ and $\alpha$ are not well determined from the experiment, and we adjust them here to optimize the fit to the visibility data shown in Fig.~\ref{fig:visiRate}. For the plots shown in this paper, we use
	\begin{equation}
		h =  {0.042~}{\rm \mu eV}, \quad   \quad \alpha = {47~}{\rm  ns^{-1.7}}.\label{eq:ah}
	\end{equation} 
	
	We note that if one takes the low and high frequency cutoffs of the noise spectrum to be ${0}{~\rm Hz}$ and $1/T_2^*$ with $T_2^*\simeq 1700{~\rm ns}$ respectively, this value of $\alpha$ yields
	a standard deviation of the
	detuning fluctuations 
	of ${5.7~}{\rm \mu eV}$, which compares well to the experimental estimate of ${6.4~}{\rm \mu eV}$ ~\cite{Wu2014}. \zqi{Moreover, we expect the magnetic energy difference $h_x$ to be in the range of ${0.01\sim 0.1}~{\rm \mu eV}$, since the $S-T_0$ experiment reported for the same device in Ref.~[\onlinecite{Wu2014}] indicates a value of $h\simeq {0.061}~{\rm \mu eV}$ for a slightly different magnetic field configuration. Hence, the fitting result for $h$ obtained above appears to be quite reasonable.}
	
	Using these results, we can plot the theoretically determined relaxation ration using Eq.~\eqref{eqn:transRate}. As shown in  Fig.~\ref{fig:transRate}, the transition rates are strongly peaked at the $S-T_-$ anticrossing. Moreover, we note from Eq.~\eqref{eqn:transRate}, the transition rates are large only if the $S-T_-$ anticrossing is not too far from the charge anticrossing. 
	
	\subsection{Oscillation Decay\label{subsec:dephasing}}
	\begin{figure}
		\includegraphics[scale = 0.3]{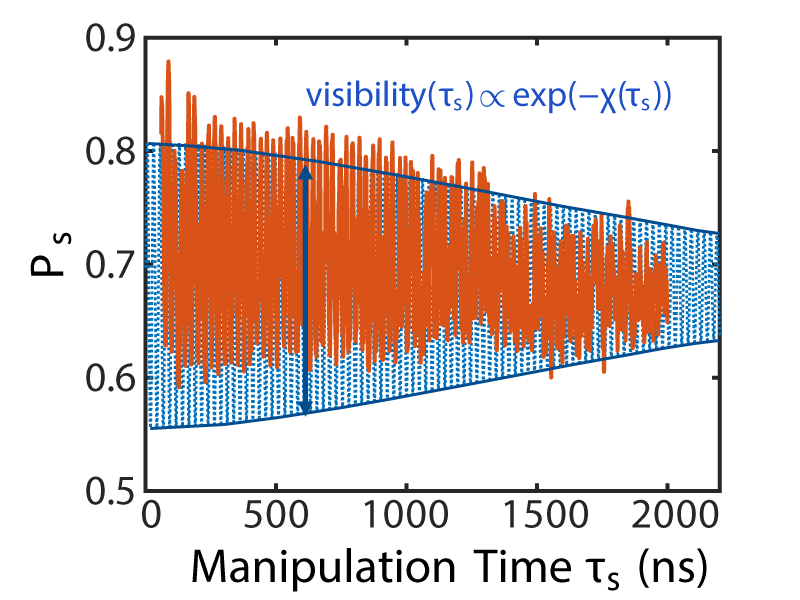}			
		\caption{ (Color online) Return probability $P_s$ measured in the experiment (red solid line) as a function of manipulation time $\tau_s$ (this is the same data as Fig.~\ref{fig:visiRate}). A Gaussian fit to the oscillation envelope yields a decoherence time of $\sim 1.7~{\rm \mu s}$~\cite{Wu2014}. The maximum visibility, or oscillation amplitude, is about $0.24$. \zqi{The blue dashed line corresponds to our theoretical prediction for the LZS oscillations, obtained using the same parameters as Fig.~\ref{fig:visiRate} at a ramp rate of $v = {4.4}~{\rm \mu eV/s}$, which is the same as the experimental ramp rate.} The blue solid lines indicate the envelope of the theoretical LZS oscillations, whose amplitude decays as $\exp(-\chi(\tau_s))$, as discussed in the main text.} 		
		\label{fig:PsTau_exp}
	\end{figure}

	To account for dephasing during the manipulation period $\tau_s$, which occurs at large detuning $\varepsilon_2\gg t_c$ far away from charge degeneracy point, we follow Refs.~\onlinecite{Makhlin2004, Astafiev2004}. The $\hat{\sigma}_z$ term in Eq.~\eqref{eqn:V2} gives rise to fluctuations of the phase difference between the qubit eigenenergy states.  The amplitude of this term  is largest when the detuning is to the left of the magnetic anticrossing, where both $\phi\simeq \pi$ and $\theta\simeq \pi$.  However, this part of the system evolution does not influence the LZS interference pattern \cite{Petta2005}, since during the forward part of the process, the system remains in the ground state, while for the reverse process, a projective measurement to the ground state is performed. Here we discuss dephasing produced by the charge noise at large positive detuning, which is far to the right of the magnetic anticrossing. This part of the cycle dominates the dephasing because the system is held at large detuning $\varepsilon_2$ for a long manipulation time, $\tau_s$.

	The phase difference $\delta\varphi$ accumulated due to fluctuations of the detuning $\varepsilon$ is
		
	\begin{equation}
		\begin{aligned}
		\delta \varphi(\tau_s)
		& = Z|_{\varepsilon = \varepsilon_2}\int_{0}^{\tau_s}\dif \tau \delta \varepsilon(\tau) \\
		& =  Z|_{\varepsilon = \varepsilon_2}\left[\dfrac{\sin(\omega \tau_s)}{\omega}\xi_{\omega}^x -\dfrac{1-\cos(\omega \tau_s)}{\omega}\xi_{\omega}^y\right],
		\end{aligned}
	\end{equation}
	where, $\delta \varepsilon(\tau) = \xi_{\omega}^x\cos(\omega t)+ \xi_{\omega}^y\sin(\omega t)$, and $\xi_{\omega}^x$ and $\xi_{\omega}^y$ are the two components of the fluctuating Gaussian fields. 
	We can compute the average of $\rho_{01}$ with respect to fluctuation of $\delta\varepsilon$ in Eq.~\eqref{subeq:off-diagonalrho} yielding $\rho_{01}(\tau_r+\tau_s) = \rho_{01}(\tau_r) \exp (-\chi(\tau_s)) \exp(-\Gamma(\varepsilon_2) \tau_s)$ (see also Refs.~[\onlinecite{Makhlin2004, Astafiev2004}]), where $\exp(-\chi(\tau_s)) = \langle \exp(-i\delta \varphi(\tau_s))\rangle$ is evaluated by 
		
		
	
	\begin{equation}
		\chi(\tau_s) =  Z^2|_{\varepsilon= \varepsilon_2}\int \dif \omega \dfrac{S(\omega)}{4}\left(\dfrac{\sin(\omega \tau_s/2)}{\omega /2}\right)^2.
		\label{eqn:dephasing}
	\end{equation}
	
	Taking the noise spectral power to be $S(\omega) = \alpha\omega^{-0.7}$ \cite{Dial2013} with parameter $\alpha = {47~}{\rm  ns^{-1.7}}$ obtained by fitting  Fig.~\ref{fig:visiRate}, one is able to calculate the time-dependence of the LZS oscillations. 
	Multiplying the factor $\exp(-\chi(\tau_s))$ by the amplitude of the return probability $P_s$ obtained in Eqs.~\eqref{eqn:differentialequations} automatically takes into account both of the dephasing factors $\Gamma$ and $\chi$, yielding a theoretical prediction for LSZ oscillations under charge noise, that is in good agreement with the experiment, as shown in Fig.~\ref{fig:PsTau_exp}. We note that the parameters used to generate these oscillations are the same as those used in Fig.~\ref{fig:visiRate}.

	\begin{figure}
		\begin{minipage}{0.22\textwidth}
			\subfigure{
				\includegraphics[scale = 0.3]{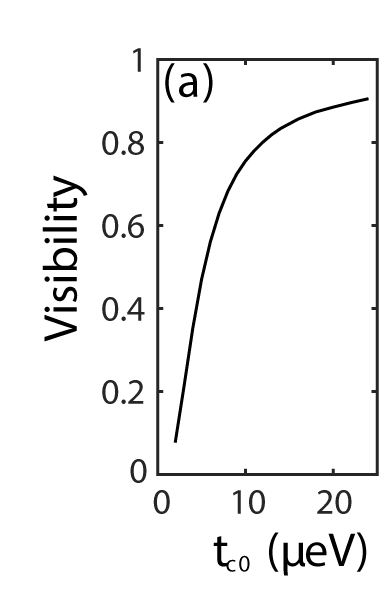}
				\label{fig:visiMaxTc0}}
		\end{minipage}
		\begin{minipage}{0.25\textwidth}
			\subfigure{
				\includegraphics[scale = 0.3]{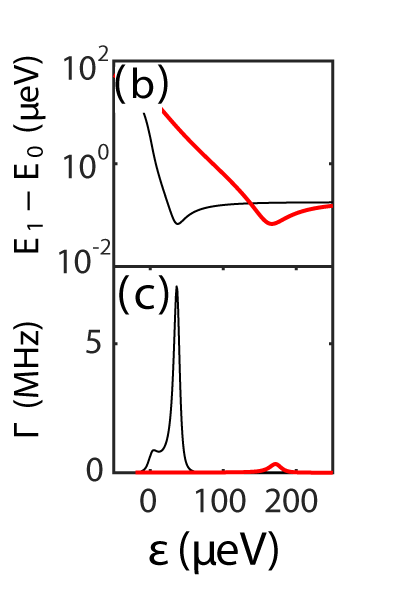}
				\label{fig:transRate}}
		\end{minipage}
		\caption{
			(a) Visibility of LZS oscillations at the optimum ramp rate as a function of tunnel coupling $t_{c0}$. (b) Semilog plot of the energy difference between the ground state $\ket{0}$ and first excited state $\ket{1}$ as a function of detuning; the dips of these curves occur at the magnetic anticrossings. (c) Transition rates $\Gamma $ as a function of detuning. The thin black lines in (b) and (c) are obtained for the tunnel coupling $t_{c0} = 3.4~{\rm \mu eV}$ while the thick red lines are obtained for $t_{c0} = 20~{\rm \mu eV}$.  Increasing the tunnel coupling can be seen to move the magnetic anticrossing farther from the charge anticrossing, causing a decrease in the transition rates.}
	\end{figure}
	Figure~\ref{fig:visiRate} and \ref{fig:PsTau_exp} indicate that our numerical results for the oscillation visibility as a function of ramp rate including only charge noise \zqi{agree well} with the experimental data; the visibility as a function of ramp rate, the coherence time of the oscillations, and the long-time limit of the decay curve can all be described by a single set of the parameter
	values  $\alpha$ and $h$ given in Eq.~\eqref{eq:ah}.
	The system exhibits both low visibility and long coherence times because the transition rates induced by the charge noise depend strongly on the detuning $\varepsilon$. While these transitions suppress the visibility  at the magnetic anticrossing, their effect is very weak at the large detuning values where phase is accumulated, so the coherence time is affected mainly by dephasing, caused by the $\hat{\sigma}_z$ term in Eq.~\eqref{eqn:V2}. 
	
	Our theory predicts that the visibility of the LZS oscillations can be increased by increasing the tunnel coupling， as shown in Fig.~\ref{fig:visiMaxTc0}. This improvement occurs because increasing the tunnel coupling increases the difference in detuning between the charge and magnetic anticrossings; when the anticrossings are well-separated, $\sin ^4(\theta/2)$ and $\sin ^2(\phi)$ in Eq.~\eqref{eqn:transRate} cannot be large simultaneously.

	\section{Discussion}
	We have shown that charge noise, which causes the detuning parameter to fluctuate in a double quantum dot, can give rise to low visibility of LZS oscillations even when the decoherence time is very long.  The key physics is that decoherence processes are greatly enhanced at the $S-T_-$ anticrossing, which decreases the visibility, but are suppressed at large detuning, leading to a long decoherence time. Our numerical results agree well with the experimental data using fitting parameters that \zqi{not only agree with the experimental estimates, but also give the same dephasing time as the experiment.}
	
	We have shown that the experimental results agree quantitatively with a theory that includes
	only charge noise, with no dephasing from nuclear spins.
	Moreover, the measured decay time of the LZS oscillations, $1.7~\mu$s, is much longer than the
	decoherence time due to nuclear spins of $0.25~\mu$s measured for an $S-T_0$ qubit in the same device~\cite{Wu2014}.
	This apparent lack of dephasing from nuclear spins is striking, and could be evidence that the experimental procedure causes an in essentially complete dynamic nuclear polarization~\cite{Reilly2008,Chekhovich2013}. \zqi{Such behavior is advantageous for quantum computing; however it is not guaranteed to occur in all experiments. More generally, we would expect to dephasing from both the nuclear fluctuation and charge noise in experiments.}
	
	In conclusion, we expect that the noise models described here could also apply to other types of qubits that exhibit low visibility and long decoherence time\cite{Harvey-Collard2015}. We also expect these results to be of interest to other experimentalists because they predict that the visibility of LZS oscillations can be increased substantially by increasing the interdot tunnel coupling.
	
	\begin{acknowledgments}
	This work was supported in part by Army Research Office Grants W911NF-17-1-0274 and W911NF-14-1-0080, National Science
	Foundation (NSF) Grants DMR 0955500, DMR-1206915 and PHY-1104660, and the Department of Defense. The authors would also like to acknowledge support from the Vannevar Bush Faculty Fellowship program sponsored by the Basic Research Office of the Assistant Secretary of Defense for Research and Engineering and funded by the Office of Naval Research through grant no.\ N00014-15-1-0029.  
	The views and conclusions contained in this document are
	those of the authors and should not be interpreted as representing the official policies, either expressly or implied,
	of the US Government. The U.S. Government is
		authorized to reproduce and distribute reprints for Government
		purposes notwithstanding any copyright notation herein. Development and maintenance of growth facilities used for fabricating samples is
	supported by Department of Energy Grant DE-FG02-03ER46028, and nanopatterning made use of NSF-supported
	shared facilities (DMR-1121288).
	Sandia National Laboratories is a multi-program laboratory managed and operated by Sandia Corporation, a wholly owned subsidiary of Lockheed Martin Corporation, for the U.S. Department of Energy's National Nuclear Security Administration under contract DE-AC04-94AL85000.
	\end{acknowledgments}
	
	\appendix*
		
		\begin{widetext}			
		\section{Reduction of the five--level system to a qubit subspace}

		In this Appendix, we show that the experimental system is well-described by a
		two-level Hamiltonian by explicitly reducing the five level system to a two-dimensional	subspace.

		The full five level system, shown in Fig.~\ref{fig:energyDiagram} of the main text, is described by the Hamiltonian

			\begin{equation} 
			\hat{H}_0^{(0)}
			=
			\left(\begin{matrix}
			\dfrac{\varepsilon}{2} & t_c & 0 & 0 & 0 \\ 
			t_c & -\dfrac{\varepsilon}{2} & \dfrac{h_x}{\sqrt{2}} & h_z & -\dfrac{h_x}{\sqrt{2}} \\ 
			0 & \dfrac{h_x}{\sqrt{2}} & -\dfrac{\varepsilon}{2}-Ez & 0 & 0 \\ 
			0 & h_z & 0 & -\dfrac{\varepsilon}{2} & 0 \\ 
			0 & -\dfrac{h_x}{\sqrt{2}} & 0 & 0 & -\dfrac{\varepsilon}{2}+E_z
			\end{matrix} \right)
			,
			\label{eqn:originalHamiltonian}
			\end{equation}

		where we use the standard basis states, given by the $(2, 0)$ singlet, the $(1, 1)$ singlet and the $T_-$, $T_0$ and $T_+$ $(1,1)$ triplets.
		Here, $\varepsilon$ is the detuning, $E_z = g\mu_B B$ is the Zeeman splitting of triplet states produced by the average  magnetic field at two dots, while off-diagonal matrix elements $h_x = g\mu_B \delta B_x$ ($h_z = g\mu_B \delta B_z$) originate from the gradient of magnetic field between the dots in the direction perpendicular to (along) the averaged field $B$ and $t_c(\varepsilon) = t_{c0} \exp(-\varepsilon/\varepsilon_0)$ is the tunnel coupling, which depends on $\varepsilon$, where $\varepsilon_0 = 125 ~{\rm \mu eV}$ is obtained from the experiment~\cite{Wu2014}.
		In this theoretical model, as in the experiment, we assume the following energy scale hierarchy:  $t_c\gg E_z\gg h_x$.  
		
		We first apply a unitary transformation $W$ defined by the matrix  
		
		\begin{equation}
		\label{eq:U1}
		\hat W = \begin{pmatrix}\displaystyle
		e^{i\sigma_y\frac{\theta}{2}} & \mathbf{0}_{2\times3}\\ \mathbf{0}_{3\times2} & \mathbf{1}_{3\times 3}
		\end{pmatrix},\quad 
		\theta = \arccos \left( \frac{\varepsilon}{2E_S}\right),
		\end{equation} 
		where  $E_S = \sqrt{(\varepsilon/2)^2+t_c^2}$. 
		This transformation diagonalizes the Hamiltonian~\eqref{eqn:originalHamiltonian} in the singlet subspace. 
		It is important to note here that the detuning parameter is a function of time, as shown in Fig.~\ref{fig:pulse_shape}. In the transformed frame, we work in the adiabatic,  time-dependent basis. The Hamiltonian then becomes:
		\begin{equation}
		\hat H_0^{(1)} = \hat W \hat H_0^{(0)} \hat{W}^\dagger - i\hat W  \dot{\hat{W}}^\dagger
		= 
		\left(\begin{matrix}
		E_S & i\dot\theta & \dfrac{h_x}{\sqrt{2}}\sin\dfrac{\theta}{2} & h_z\sin\dfrac{\theta}{2} & -\dfrac{h_x}{\sqrt{2}}\sin\dfrac{\theta}{2} \\ 
		i\dot\theta & -E_S & \dfrac{h_x}{\sqrt{2}}\cos\dfrac{\theta}{2} & h_z\cos\dfrac{\theta}{2} & -\dfrac{h_x}{\sqrt{2}}\cos\dfrac{\theta}{2} \\ 
		\dfrac{h_x}{\sqrt{2}}\sin\dfrac{\theta}{2} & \dfrac{h_x}{\sqrt{2}}\cos\dfrac{\theta}{2} & -\dfrac{\varepsilon}{2}-E_z & 0 & 0 \\ 
		h_z\sin\dfrac{\theta}{2} & h_z\cos\dfrac{\theta}{2} & 0 & -\dfrac{\varepsilon}{2} & 0 \\ 
		-\dfrac{h_x}{\sqrt{2}}\sin\dfrac{\theta}{2} & -\dfrac{h_x}{\sqrt{2}}\cos\dfrac{\theta}{2} & 0 & 0 & -\dfrac{\varepsilon}{2}+E_z
		\end{matrix} \right),
		\label{eq:H01}
		\end{equation}
		
			\begin{figure}
			\begin{minipage}{0.43\textwidth}
				\subfigure{%
					\includegraphics[scale = 0.3]{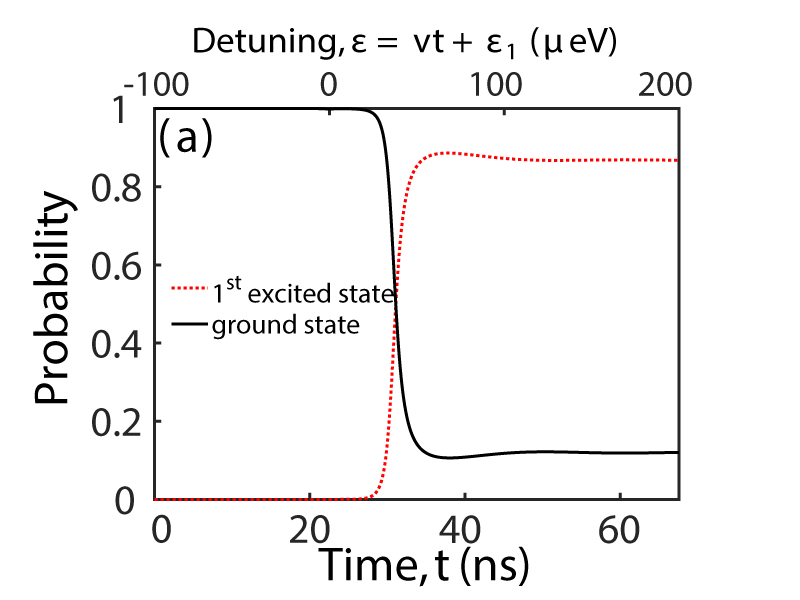}
					\label{fig:probFirstTwo}}
			\end{minipage}
			\begin{minipage}{0.43\textwidth}
				\subfigure{%
					\includegraphics[scale = 0.3]{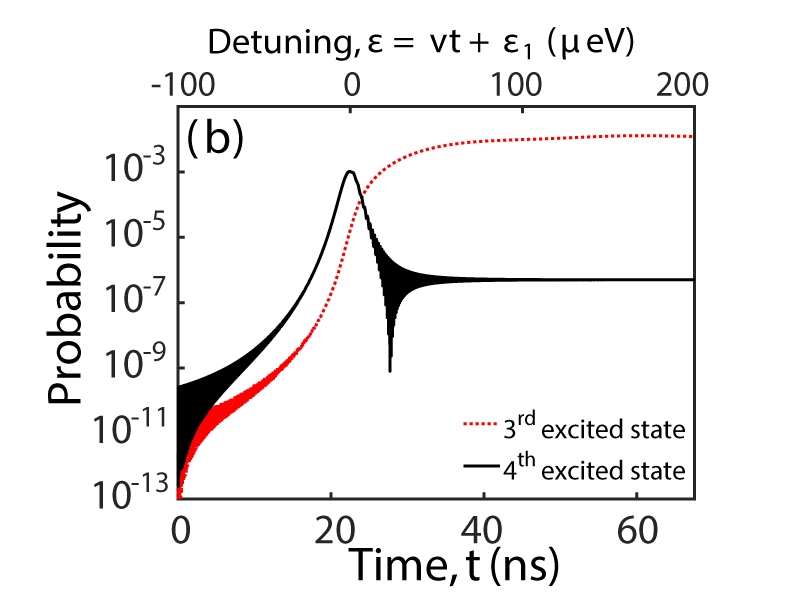}
					\label{fig:probHigherExcited}}
			\end{minipage}
			\caption{(a) shows the probability of the ground state $\ket{0}$ and the first excited state $\ket{1}$ of a five-level system through a one-directional ramp, starting from the ground state at detuning $\varepsilon_1 = -100 ~{\rm \mu eV}$. The ramp rate is $v = \dif \varepsilon/\dif t = 4.4 ~{\rm \mu eV}$. Parameters $t_{c0} = {3.4}{~\rm \mu eV}$, $h =  {0.042~}{\rm \mu eV}$ and $g \mu_B B = {0.17}{~\rm\mu eV}$ are the same as for Fig.~\ref{fig:visiRate} in the main text. Charge noise is not included in this calculation. The occupations of the other three states are too small to be visible in (a). (b) shwows the probability of the third and fourth excited states. The second excited state is not coupled to the other states given that $h_z = 0$, and the probability being in this level remains zero throughout the ramp. Because only two of the energy levels have significant occupation at any time during the evolution, the dynamics can be described using the two-state Hamiltonian  (Eq.~\eqref{eq:H1st} in the main text).}
		\end{figure}
		
	\end{widetext}

		where  the term $\hat W  \dot{\hat{W}}^\dagger$ originates from time dependence of the transformation $\hat W$ and results in $\dot{\theta}$ terms in $\hat H_0^{(1)}$.  
		The time derivative of the transformation angle $\theta$ is

		\begin{equation}
		\dot{\theta} = -2
		\dfrac{t_c(\varepsilon) -\varepsilon (\partial t_c(\varepsilon) /\partial \varepsilon ) }
		{\varepsilon^2+4t_c^2(\varepsilon)}v,
		\end{equation}
		where $v = \dif \varepsilon /\dif t$. Below, we assume the longitudinal field $h_z = 0$, in which case the $T_0$ state decouples from the other four states. We define the qubit states $\ket{0}$ as the ground state and $\ket{1}$ as the lowest excited state of the Hamiltonian. In the limit of $h_x\rightarrow0$, these eigenvectors are the low energy singlet and triplet $T_-$ states with energies $-E_S$ and $E_{T_-} =-\varepsilon/2-E_z$, respectively. 
		The minimal energy gap between $\ket{1}$ and $\ket{0}$ is $h = h_x\sqrt{2}\cos(\theta(\varepsilon)/2)$ evaluated at the detuning $\varepsilon^*$ such that $E_S(\varepsilon^*) = \varepsilon^*/2+E_z$.
		This small energy splitting makes it possible to observe the LZS oscillations at relatively low detuning ramp rates $v\lesssim 10~{\rm \mu eV/ns}$. These values of the ramp rate are too small to cause transitions out of the qubit subspace, justifying our approximation of a two-level system. 
		
		We have tested our approximation of a two-level system by performing simulations on a full five-level system, obtaining the results shown in Fig.~\ref{fig:probFirstTwo} and \ref{fig:probHigherExcited}. The population of the third excited state (the $T_+$ state) is found to be of the order of $10^{-2}$, and the fourth excited state (the higher energy singlet) is of the order of  $10^{-5}$.
		The simulations begin in the initial $(2,0)$ singlet state, which corresponds to the lowest energy state at $\varepsilon_1=-100\mu$eV. We also assume a ramp rate of $v = (d\varepsilon/dt) =  4.4 ~\rm\mu eV/ns$. 

	\end{document}